\documentstyle[aps,amsfonts,multicol,prl,prbbib,epsfig]{revtex}

\begin{document}

\newcommand {\pd}[2]{ \frac{\partial #1}{\partial #2}}
\newcommand{\fd}[2]{ \frac{\delta #1}{\delta #2}} \newcommand
{\bv}[1]{ \mathbf{#1} } \newcommand {\mchi}{\bar{\chi}} \newcommand
{\hv}[1]{\hat{\bv #1}} \newcommand
{\gf}[2]{\Gamma\!\left(\frac{#1}{#2}\right)}

\author{ A.N. Morozov$^{1,2}$ \and J.G.E.M. Fraaije$^2$}

\address{$^1$Faculty of Mathematics and Natural Sciences, University
of Groningen, \\ Nijenborgh 4, 9747 AG Groningen, The Netherlands,
email: a.n.morozov@chem.rug.nl}

\address{$^2$Soft Condensed Matter group, LIC, Leiden University, PO
Box 9502, 2300 RA Leiden, The Netherlands}

\title{Orientations of the lamellar phase of block copolymer melts
under oscillatory shear flow}

\maketitle

\begin{abstract} We develop a theory to describe the reorientation
phenomena in the lamellar phase of block copolymer melt under
reciprocating shear flow. We show that similar to the steady-shear,
the oscillating flow anisotropically suppresses fluctuations and gives
rise to the $\parallel\rightarrow\perp$ transition. The experimentally
observed high-frequency reverse transition is explained in terms of
interaction between the melt and the shear-cell walls.
\end{abstract}

\begin{multicols}{2}

The behaviour of the lamellar phase (a stripped pattern) of block
copolymer melts under oscillatory shear flow have attracted attention
of numerous experimental studies
\cite{koppi92,koppi93,zhang95,patel95,wiesner99,hamley:book}.  Shear
flow is known to influence the order-disorder transition (ODT)
temperature and the orientation of the lamellae with respect to the
shear geometry. Thus, in the vicinity of ODT at low frequencies the
lamellae orient with their normal parallel to the shear gradient (the
parallel orientation), while at higher frequencies their normal is
perpendicular to the velocity and the gradient directions (the
perpendicular orientation). Further increase of frequency results in
reappearance of the parallel orientation \cite{wiesner99}.  In this
Letter we propose an explanation of this orientation behaviour which
is usually referred to as a double-flip phenomena.

Earlier theories, which deal with steady shear, emphasize the role of
compositional fluctuations
\cite{onuki_kawasaki,cates_milner,j_rheol}. The stable orientation is
seen as a result of interaction between the shear flow and the
fluctuation spectra. In equilibrium fluctuations destroy the
long-range correlations and therefore lower the ODT temperature with
respect to its mean-field value. Imposition of shear breaks the
rotational symmetry and anisotropically suppresses fluctuations. The
direction of the strongest suppression will have the highest ODT
temperature and the corresponding orientation of lamellae will be
selected. We will show that the selected orientation depends on the
amplitude and frequency of the flow. We base our analysis on the
Fokker-Planck equation for the probability density $P[\phi]$
\begin{eqnarray}
\label{FP} \pd{P}{t}[\phi,t]&=&\int_{k}\fd{ }{\phi_{\bv k}} \biggl[\mu
\left( \fd{ }{\phi_{-{\bv k}}}+ \fd{H[\phi]}{\phi_{-{\bv k}}} \right)
\nonumber \\ &&-A \omega \cos{\omega t}\,k_{x}\pd{}{k_y}\phi_{\bv{k}}
\biggr]P[\phi,t]\,.
\end{eqnarray} Here $\phi_{\bv k}$ is a fluctuating scalar field
described by the Brazovskii Hamiltonian \cite{Fredrickson:1987}
\begin{eqnarray}
\label{ham} & & \qquad \qquad H[\phi]=\frac{1}{2}\int_{k} [\tau
+(k-k_{0})^2]\phi_{\bv{k}}\phi_{-\bv{k}}\\
&&+\frac{1}{4!}\int_{k_1}\int_{k_2}\int_{k_3}\int_{k_4}
\lambda({\bv{k}}_1, {\bv{k}}_2,{\bv{k}}_3,{\bv{k}}_4) \phi_{{\bv
k}_1}\phi_{{\bv k}_2}\phi_{{\bv k}_3}\phi_{{\bv k}_4} \nonumber \,,
\end{eqnarray} $\tau$ is a temperature-controlling parameter,
$k_0^{-1}$ is an intrinsic length-scale of the block-copolymer melt
arising from the interplay of interactions and the chain connectivity,
$\mu$ is an Onsager coefficient which is approximated by
$\mu=\mu(k_0)$ and assumed to be frequency independent
\cite{Binder:1983,Fredrickson:1986,muthu_shear}. The last term in
Eq.(\ref{FP}) describes a coupling between the shear flow ${\bv v}=A
\, \omega \cos{\omega t} \,y\, {\bv e}_x$ and the gradient of the
order parameter. Here we assume that this form of flow is valid for
all $\omega$ and $A$.

The Fokker-Planck equation (\ref{FP}) generates the equations for the
amplitude of the average order-parameter profile $\langle \phi_{\bv k}
\rangle=a (\delta_{{\bv k} , k_0 {\bv n}}+\delta_{{\bv k}, -k_0 {\bv
n}})$ oriented along the unit vector ${\bv n}$
\begin{eqnarray}
\label{amp} \frac{1}{\mu}\pd{a}{t}=-r({\bv
n})a+\frac{\lambda}{2}(1-\beta)a^3 \,,
\end{eqnarray} and for the structure factor $S({\bv k})=\langle
\phi_{\bv k}\phi_{-\bv k}\rangle- \langle \phi_{\bv k}\rangle
\langle\phi_{-\bv k}\rangle$
\begin{eqnarray}
\label{S} \frac{1}{2\mu}\pd{S({\bv k})}{t}-\frac{A \, \omega}{2 \mu}
\cos{\omega t} \, k_x \pd{S({\bv k})}{k_y}+S({\bv k}) S({\bv
k})_0^{-1}=1 ,
\end{eqnarray} where
\begin{eqnarray}
\label{not} &&\quad \quad \qquad S_0^{-1}({\bv k})=r(\hat{\bv
k})+(k-k_0)^2\,,\\ & &r(\hat{\bv k})\equiv r-\hat{\bv k}\cdot
\stackrel {\leftrightarrow}{\bv e} \cdot\hat{\bv k} =\tau+\lambda a^2
\left(1-\beta \left({\bv n}\cdot \hat{\bv k}
\right)^2\right)+\sigma(\hat{\bv k})\,, \nonumber \\ && \quad
\sigma(\hat{\bv k})=\frac{\lambda}{2}\int \frac{d{\bv q}}{(2\pi)^3}
S({\bv q}) \left[1-\beta \left(\hat{\bv k}\cdot\hat{\bv q} \right)^2
\right]\,. \nonumber
\end{eqnarray} The interaction \cite{comm1} between fluctuations
$\lambda({\bv k},-{\bv k},{\bv q},-{\bv q})=
\lambda\left[1-\beta(\hat{\bv k}\cdot \hat{\bv q})^2 \right]$, with
$\hat{\bv k}={\bv k}/k$, renormalizes the temperature $r(\hat{\bv k})$
and makes it anisotropic in the presence of shear.

The stability criterion for an orientation is derived from
Eq.(\ref{amp}). It has a potential form $\partial a/\partial
t=-(\mu/2)\partial \Phi(a)/\partial a$, with
\begin{eqnarray}
\label{pot} \Phi(a,{\bv n})=-\frac{1}{4}\lambda a^4(1-\beta)+2\int_0^a
da' r({\bv n})a'\,.
\end{eqnarray} Generally, the potential $\Phi$ is time-dependent. In
steady-state, however, it oscillates around some average value. In
order to simplify our analysis we coarse-grain the time-scale with the
period of oscillations and consider the time-independent version of
Eq.(\ref{not}-\ref{pot}) with $\sigma(\hat{\bv k},t)$ replaced by
$\bar{\sigma}(\hat{\bv k})= (\omega/2\pi)\int_0^{2\pi/\omega}dt
\,\sigma(\hat{\bv k},t)$. In this model the minimum of $\Phi(a,{\bv
n})$ determines the stable orientation \cite{comm2}. The potential
$\Phi$ can be viewed as a dynamical extension of the equilibrium
free-energy.  Together with the solution of Eq.(\ref{S})
\begin{eqnarray}
\label{sol} & &S({\bv k},t)=2\mu\int_0^t d\tau \exp{\biggl[-2\mu
\int_\tau^t ds \,S_0^{-1}\left({\bv k}(s)\right) \biggr]}\,, \\ &
&\quad {\bv k}(s)=(k_x,k_y+A k_x (\sin{\omega t}-\sin{\omega
s}),k_z)\,, \nonumber
\end{eqnarray} it allows us to construct the orientational phase
diagram.

First we consider the low-amplitude shear ($A\ll1$). In this case we
expand the exponent in (\ref{sol}) up to $O(A^2)$ and obtain
\begin{eqnarray}
\label{low} &&\bar{\sigma}({\bv n})=\sigma_1-\frac{3\pi \beta}{56}
\left(\frac{A \omega}{2\mu}\right)^2 \frac{\alpha^2\lambda}{r^{7/2}}
\left(n_y^2+\frac{n_z^2}{3}\right)\,, \, \omega\ll 1 \,,\nonumber \\
&&\bar{\sigma}({\bv n})=\sigma_2-\frac{2\pi \beta}{35} A^2
\frac{\alpha^2\lambda}{r^{3/2}}
\left(n_y^2+\frac{n_z^2}{3}\right)\,\,, \,\, \omega\gg 1 \,,
\end{eqnarray} where $\alpha=k_0^2/(4\pi)$ and $\sigma_1$, $\sigma_2$
absorb the orientation-independent terms. In the low-amplitude regime
the fluctuation spectra is only slightly influenced by shear
flow. Eqs.(\ref{low}) show that this regime is a perturbation of the
equilibrium state with $\bar{\sigma}=\sigma_0$. When the frequency is
low, the typical time of the critical fluctuation development is much
shorter than $\dot{\gamma}^{-1}\equiv(A \omega)^{-1}$ and the flow
simply translates fluctuations in space. At high frequencies the
life-time of the critical fluctuations exceeds the characteristic time
of the flow. However, since the amplitude of deformation is small,
fluctuations live in an averaged environment, similar to the
equilibrium one. Then the properties of the melt cannot depend on the
frequency, which is shown by Eq.(\ref{low}). To determine the stable
orientation we follow Fredrickson and note that the fluctuation
integral $\bar{\sigma}$ is smaller for the parallel orientation
($n_y=1$).  The fluctuations are weaker in this direction and
therefore we predict the parallel orientation to be stable in the
low-amplitude regime.

For the finite-amplitude shear the fluctuation integral $\bar{\sigma}$
can be evaluated with the help of Eq.(\ref{sol}) near $k_x\approx0$
\begin{eqnarray}
\label{high} & &\bar{\sigma}({\bv n})=c_1(\alpha
\lambda)^{2/3}\left(\frac{D_*}{A\omega}
\right)^{1/3}\left[1-\frac{\beta} {7}(2 n_y^2+3 n_z^2)
\right]\,,\,\omega\ll1 \nonumber \\ &&\bar{\sigma}({\bv
n})=c_2\frac{A_*}{A}\biggl[\ln^2{\epsilon}-\pi^2 \\
&&\qquad\quad-\beta\left(4n_y^2
\ln{\frac{\epsilon}{e^4}}+n_z^2(16-\pi^2+\ln^2{\frac{\epsilon}{e^4}})
\right) \biggr] \,,\,\omega\gg1 \,, \nonumber \\ && \quad D_*=\mu
\lambda\sqrt{\alpha} \quad,\quad A_*=\lambda\sqrt{\alpha} \quad,\quad
\epsilon=\frac{32\sqrt{3}A^2 k_0^2}{r({\bv n})} \,,\nonumber \\ &&
\quad \quad c_1=\frac{\Gamma\left(\frac{1}{6}
\right)3^{1/6}}{\pi^{2/3}} \qquad,\qquad c_2=\frac{\sqrt{2}}{16
\pi^{3/2}3^{1/4}}\,. \nonumber
\end{eqnarray} This regime no longer resembles the equilibrium
state. Indeed, in equilibrium \cite{Fredrickson:1987} as well as under
low-amplitude shear the spinodal temperature, given by $r({\bv
n})|_{a=0}=0$, is suppressed by fluctuations to $\tau_s=-\infty$.
Unlikely, from Eq.(\ref{high}) $\tau_s({\bv n})=-\bar{\sigma}({\bv
n})$ asymptotically approaches zero as $A\rightarrow\infty$. We
conclude that the flow strongly suppresses fluctuations and restores
the mean-field behaviour. We also see that $\tau_s(n_z)>\tau_s(n_y)$
and therefore we expect the perpendicular orientation to appear below
the spinodal. Analysis of the potential $\Phi$ shows that this
orientation will persist for lower temperatures.

The transition from the parallel to perpendicular orientations can be
located with the help of the method from Ref.\cite{ik_walls}. We
interpolate $\bar{\sigma}$ in-between the $A\rightarrow0$ and
$A\rightarrow\infty$ regimes and solve the equation
$\tau_s(n_y)=\tau_s(n_z)$ to obtain the transition line in the
$A-\omega$ plane
\begin{equation}
\label{line} A\;\omega\approx 10^3\mu\; k_0^2 \sim N^{-3} \quad,\quad
\omega\ll1 \, ,
\end{equation} where $N$ is a number of monomers in a polymer chain.
For high frequencies $\bar{\sigma}$ is independent of $\omega$ and we
predict the transition line to be given by $A=const$. Recent
experimental work\cite{wiesner99} argues that at low frequencies the
transition between the orientations occurs at $A\sim\omega^{-1}$,
while at higher frequencies the transition line starts to level
off. This is in a full agreement with our predictions.

Finally, we want to discuss the reappearance of the parallel
orientation at high frequencies. This behaviour cannot arise from the
flow--fluctuation interaction discussed above. Some authors argue that
the assumption of slow flow in Eq.(\ref{FP}) is responsible for the
failure of the theory to predict the parallel orientation at high
frequencies \cite{Bates_Fredrickson:dynamics,hamley:book}. We,
however, support a different opinion. Balzara {\it et. al}
noticed\cite{balsara94} that in equilibrium the walls of shear cell
induce the parallel alignment through the whole 0.5-mm sample, while
Laurer {\it et. al} observed \cite{laurer} that under shear there is
always a near-surface layer of the parallel lamellae independently of
the bulk orientation. Therefore we propose that the high-frequency
parallel orientation of the lamellae is caused by interactions of the
shear-cell walls with the melt. Fredrickson has shown \cite{fr_surf}
that in equilibrium this interaction will lead to the parallel
alignment. Recently we discussed this effect for a steady-shear
\cite{ik_walls} and showed that in the presence of this interaction
the stable orientation is given by the minimum of $\Phi'=\Phi-2\eta a
\delta_{n_y,1}$, where $\eta$ is proportional to the Flory-Huggins
strength of interaction between the walls and melt. Minimization of
the modified potential $\Phi'$ gives for the
$\perp\rightarrow\parallel$ transition temperature
\begin{equation}
\tau_1=-\bar{\sigma}(n_y)-\frac{\left[\bar{\sigma}(n_y)-\bar{\sigma}(n_z)\right]^4}{8
\eta^2 \lambda (1-\beta)} \,,
\end{equation} with $\bar{\sigma}({\bv n})$ from Eq.(\ref{high}). If
we fix temperature, the line of the second transition will be again
given by $A\sim\omega^{-1}$, $\omega\ll1$ and $A=const$, $\omega\gg1$
with coefficients depending on $\eta$. This is in a qualitative
agreement with experiments \cite{wiesner99}. A detailed comparison is
impossible because of the lack of experimental data. Our assumption
can be verified by performing measurement in various material shear
cells.

We summarize our results in orientational diagram (Fig.1). The
low-frequency regime resembles the steady-shear behaviour. The
corresponding expressions for $\bar{\sigma}$
(Eqs.(\ref{low},\ref{high})) are similar to those for the steady-shear
\cite{cates_milner,j_rheol} with an effective shear rate
$\dot{\gamma}=A\,\omega$. The high-frequency part of the diagram
fundamentally differs from the steady-shear.  When the frequency
exceeds some critical value, which is of order of the relaxation time
for the critical fluctuations, the further increase of frequency does
not change the behaviour of the system. Therefore, in the
high-frequency limit $\bar{\sigma}({\bv n})$ together with the
spinodal temperature $\tau_s({\bv n})$ appears to be independent of
$\omega$.

The first transition from the parallel to perpendicular orientation
corresponds to a change in character of the flow--fluctuation
interaction. This change is associated with a strong suppression of
fluctuation and a cross-over from the fluctuation to mean-field type
of behaviour. Our estimate for the critical effective shear rate
(Eq.(\ref{line})) shows that
\cite{Binder:1983,Fredrickson:1986,muthu_shear,ik_walls}
$\dot{\gamma}_c\sim N^{-3}$.  When $N\rightarrow\infty$, the
fluctuation region disappears \cite{Fredrickson:1987} and
$\dot{\gamma}_c\rightarrow0$.

To explain the second transition we make use of recent experiments
\cite{balsara94,laurer} and argue that the high-frequency parallel
orientation is stabilized by the preferable interaction of the
shear-cell walls with on of the components of the melt. Validity of
this hypothesis requires further experimental studies.

At the end we want to emphasize that the presented picture is
applicable for any system (polymers, surfactants, microemulsions)
described by the Brazovskii Hamiltonian (Eq.(\ref{ham})).

The authors are grateful to Boele Braaksma for his help with elliptic
and hypergeometric functions.

%\bibliographystyle{prsty}
%\bibliography{my}

\begin{figure}
\begin{picture}(0,150) \put(0,0){\epsfig{file=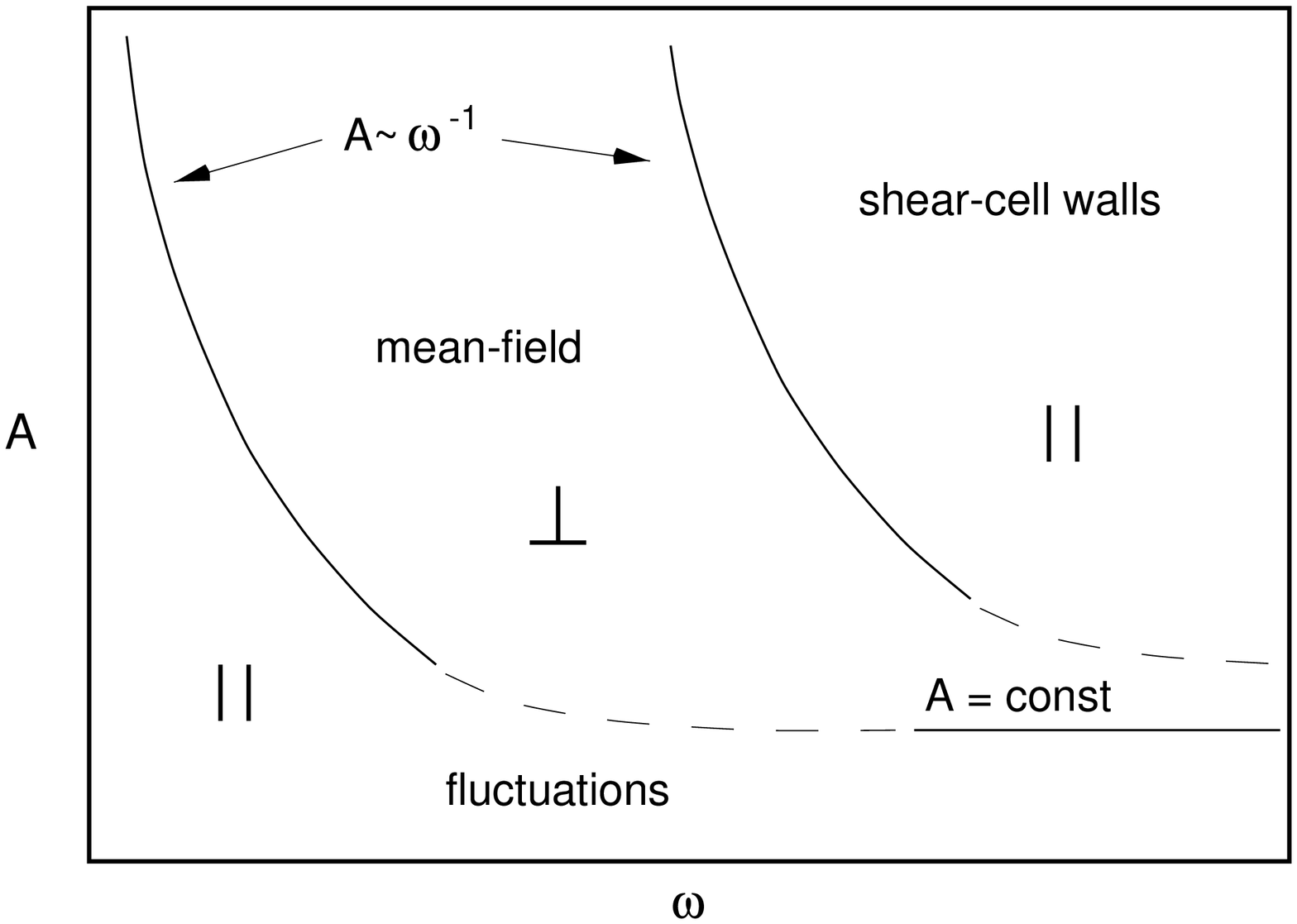,width=7cm}}
\end{picture}
\end{figure} Fig.1. Orientational diagram.  In each region the
dominated effect is stated.

\end{multicols}

\end{document}